\documentclass[preprint,showpacs,preprintnumbers,amsmath,amssymb,nofootinbib]{revtex4}
\usepackage{bbm}
\usepackage{amsfonts}
\usepackage{booktabs}
\usepackage{mathrsfs}
\usepackage{epsfig}
\usepackage{graphicx}
\usepackage{dcolumn}
\usepackage{bm}
\usepackage{amsmath}
\usepackage{slashed}

\let\jnfont=\rm
\def\NPB#1,{{\jnfont Nucl.\ Phys.\ B }{\bf #1},}
\def\PLB#1,{{\jnfont Phys.\ Lett.\ B }{\bf #1},}
\def\EPJC#1,{{\jnfont Eur.\ Phys.\ Jour.\ C }{\bf #1},}
\def\PRD#1,{{\jnfont Phys.\ Rev.\ D }{\bf #1},}
\def\PRL#1,{{\jnfont Phys.\ Rev.\ Lett.\ }{\bf #1},}
\def\MPLA#1,{{\jnfont Mod.\ Phys.\ Lett.\ A }{\bf #1},}
\def\JPG#1,{{\jnfont J.\ Phys.\ G}{\bf #1},}
\def\CTP#1,{{\jnfont Commun.\ Theor.\ Phys.\ }{\bf #1},}
\def\ZPC#1,{{\jnfont Z.\ Phys.\ C }{\bf #1},}
\def\JHEP#1,{{\jnfont JHEP \ }{\bf #1},}
\def\Rv{\not{\hbox{\kern-1pt $R$}}}
\def\p{\not{\hbox{\kern-3pt $p$}}}

\begin{document}

\title{Next-to-leading order QCD corrections to $HZW^{\pm}$ production \\ at 14 TeV LHC}

\author{Ning Liu$^{1,2}$, Jie Ren$^{1,3}$, Bingfang Yang$^{4}$
        \\~ \vspace*{-0.3cm} }
\affiliation{ $^1$ Physics Department, Henan Normal University, Xinxiang 453007, China\\
$^2$ ARC Centre of Excellence for Particle Physics at the Terascale, School of Physics,
The University of Sydney, NSW 2006, Australia\\
$^3$ State Key Laboratory of Theoretical Physics, Institute of Theoretical Physics,
 Academia Sinica, Beijing 100190, China\\
$^4$ Basic Teaching Department, Jiaozuo University, Jiaozuo 454000, China \vspace*{1.5cm}}

\begin{abstract}
Since the precise study of Higgs gauge couplings is important to test the Standard Model (SM), we calculate the complete next-to-leading order QCD(NLO QCD) correction to the $pp \to HZW^{\pm}$ production in the SM at 14 TeV LHC. Our results show that the NLO QCD correction can enhance the leading-order cross section of $pp \to HZW^{\pm}$ by $45\%$, when $ m_H $ = 125.3 GeV. We also study the dependence of the LO and NLO corrected cross sections on the renormalization and factorization scale $\mu$. Besides, due to the unbalance of parton distribution functions, we investigate the charge asymmetry of $W^{\pm}$ in the production of $pp\to HZW^{\pm}$, which can reach $32.94\%$ for $\mu=(m_H+m_Z+m_W)/2$ at 14 TeV LHC.
\end{abstract}
\maketitle

\section{ Introduction }
On 4th July 2012, the ATLAS and CMS collaborations have announced the observation of a Higgs boson with a mass around 125 GeV at the LHC \cite{ATLAS,CMS}. This discovery implies a great success of the SM and opens an era of the precise study of Higgs boson. Up to now, although most measurements of the Higgs boson properties are consistent with the SM predictions\cite{Higgs}, there are still rooms for the new physics that contributes to effective couplings to gluons and photons \cite{chgg}. Since the interactions between Higgs boson and gauge bosons are sensitive to the new physics, any modifications of the Higgs gauge couplings will lead to the deviation of the SM predictions from the Higgs data\cite{chvv}. Besides, the Higgs gauge couplings are strongly related to the anomalous triple vector bosons vertex\cite{cvvv}. Thus, the precise study of Higgs gauge couplings is important for testing the SM and searching for new physics.

There have been many works devoted to investigating the Higgs gauge couplings from the processes of $ pp \to ZH/W^{\pm}H$, which have been calculated to the next-to-leading order(NLO) in the SM at the LHC \cite{ppHZ/HW}. In addition, the production of Higgs boson associated with dibosons can also be used to probe the Higgs gauge couplings \cite{ppHVV}, such as $pp \to HW^+W^-, HZW^{\pm}, HZZ, HZ\gamma, HW^{\pm}\gamma$. Among these processes, only $pp \to HW^+W^-, HW^{\pm}\gamma$ have been recently calculated with NLO QCD corrections in Ref.\cite{song}. Due to the gauge symmetry, the ratio of tree-level production rates of $\sigma(HWW):\sigma(HZW):\sigma(HZZ)$ is predicted to be $4:2:1$ in the SM at very large energy scale. The measurement of this ratio can be served as a good test of the couplings between Higgs and a vector boson pair.

In this paper, we calculate the next-to-leading order QCD corrections to the production process of $pp \to HZW^{\pm}$ in the SM at 14 TeV LHC. Due to the trilepton signals and a reconstructed Higgs mass in the final states, the process of $pp \to HZW^{\pm}$ in the SM may be detected at the LHC. Besides, this process will be an important background of new physics signal $pp \to HW^{'\pm} \to HZW^{\pm}$ in some extensions of the SM gauge group, where the predicted extra gauge boson $W'$ only couples with the gauge bosons and Higgs boson\cite{he}. The paper is organized as follows: In Sec.II, we present the description of the NLO calculation of the process $pp \to HZW^{\pm}$. Then, we give the numerical results and discussions in Sec.III. Finally, we give a short summary in Sec.IV.

\section{description of NLO calculation of $ pp \to HZW^{+} $}

The leading order(LO) process of $p p \to HZW^{+}$ is induced by the electro-weak interaction in the SM. We denote the four-momenta of initial and final states in the process as follows:
\begin{equation}
  q(q_1) + q'(q_2) \to H(q_3) + Z(q_4) + W^+(q_5)~~(qq'=u\bar{d},u\bar{s},c\bar{d},c\bar{s}).\label{eq1}
\end{equation}
The NLO QCD corrections($\Delta\sigma_{NLO}$) to the above process can be divided into the following parts in the calculations:
\begin{itemize}
  \item the virtual corrections($\Delta\sigma^{vir}$): $qq' \to HZW^+$;
  \item the real gluon emission corrections($\Delta\sigma^{real}_{g}$): $qq' \to HZW^+g$;
  \item the real light-(anti)quark emission corrections($\Delta\sigma^{real}_{q}$): $qg \to HZW^+q'$.
\end{itemize}

We generate the one-loop amplitudes by the FeynArts-3.5\cite{feynarts}. We use the FormCalc-6.1\cite{formcalc} to simplify the amplitudes and reduce the loop functions as the definitions in Ref\cite{loop function}. We adopt the dimensional regularization to isolate all the ultraviolet divergences (UV) and infrared divergences (IR) in the virtual corrections. We remove the UV divergences by the counter terms fixed by the on-mass-shell renormalization condition\cite{on-shell}. The infrared (IR) singularities from the one-loop integral can be cancelled by adding the contributions of the real gluon emissions. We deal with the IR divergences in Feynman integrals as Ref.\cite{ellis} and implement the numerical calculations for the IR safe parts of N-point integrals as Ref.\cite{loop function}. The loop functions in the virtual corrections are numerically calculated by the modified package of LoopTools-2.2\cite{looptools}.


By using the above packages, we adopt the two cutoff phase space slicing (TCPSS) method \cite{TCPSS} to isolate the IR singularity in the real gluon emission process $ qq' \to HZW^{+}g$, where the four-momenta for the initial and final states are denoted as follows:
\begin{equation}
  q(q_1) + q' (q_2) \to H(q_3) + Z(q_4) + W^+(q_5) + g(q_6)~~(qq'=u\bar{d},u\bar{s},c\bar{d},c\bar{s}).
\end{equation}
 An arbitrary small cutoff parameter $\delta_s$ is introduced to split the phase space into soft region($E_6 \leq \delta_s \sqrt{s} / 2$) and hard region($E_6 > \delta_s \sqrt{s} / 2$). We further divide the hard part into hard collinear region $HC$ ($-\hat{t}_{16}$ or $-\hat{t}_{26} <\delta_c\hat{s}$) and hard non-collinear region $H\overline{C}$ ($-\hat{t}_{16}$ and $-\hat{t}_{26} > \delta_c\hat{s}$) by the cutoff parameter $\delta_c$, where $\hat{t}_{i6}=(p_{i}-p_6)^2, i=1,2$. The soft part and hard collinear part will be canceled with the IR divergences in the virtual corrections. So the cross section of the process of $ qq' \to HZW^{+}g$ can be decomposed as followings:
\begin{equation}
  \Delta\sigma^{real}_{g} = \Delta\sigma^{soft}_{g} + \Delta\sigma^{HC}_{g} + \Delta\sigma^{H\overline{C}}_{g}.
\end{equation}

Besides, we also consider the contribution from the real light quark emission process $ qg \to HZW^{+}q'$, where the four-momenta for the initial and final states are denoted as follows:
\begin{equation}
  q(q_1) + g (q_2) \to H(q_3) + Z(q_4) + W^+(q_5) + q'(q_6)~~(qq'=ud,us).\label{light-quark}
\end{equation}
It should be mentioned that only the initial state collinear singularities can occur in the Eq.(\ref{light-quark}), which can be absorbed
into the redefinition of the PDFs at the NLO. While, for the non-collinear part, it is computed by using the Monte Carlo technique \cite{hardMC}.So the cross section of the process of $ qg \to HZW^{+}q'$ can be divided as followings:
\begin{equation}
  \Delta\sigma^{real}_{q} = \Delta\sigma^{HC}_{q} + \Delta\sigma^{H\overline{C}}_{q}.
\end{equation}
We implement all the above calculations as in our previous works \cite{bbh,ttr,ttz} and numerically checked our results with the MadGraph5-v2 that includes the packages of MadLoop and aMC$@$NLO \cite{amcnlo,mad5} and found they are consistent in the reasonable error range.

\section{ Numerical result and discussion }

The SM parameters used in our numerical calculation are \cite{pdg}:
\begin{eqnarray}
  & \alpha_{ew} = 1/128 , \quad m_W = 80.385\mathrm{GeV} , \quad m_Z = 91.1876 \mathrm{GeV},  \quad m_t = 173.5 \mathrm{GeV}. &
\end{eqnarray}
besides, the CKM matrix elements are taken as
\begin{equation}
\begin{array}{lll}  \label{input2}
V_{CKM}=\left(
\begin{array}{ccc}
 V_{ud} & V_{us} & V_{ub}\\
 V_{cd} & V_{cs} & V_{cb}\\
 V_{td} & V_{ts} & V_{tb}\\
\end{array}\right)=
\left(
\begin{array}{ccc}
 0.97418 & 0.22577 & 0\\
 -0.22577 & 0.97418 & 0\\
 0 & 0 & 1\\
\end{array}\right).
\end{array}
\end{equation}
The Higgs mass is taken as $m_H = 125.3$ GeV from the combined results of ATLAS and CMS in Ref.\cite{giardino}. For the strong coupling constant $ \alpha_s (\mu)$, we evaluate it by the 2-loop evolution with QCD parameter $ \Lambda^{n_f = 5} = 226 $ MeV. We use CTEQ6L1 and CTEQ6M parton distribution functions (PDF) for the LO and NLO QCD computations, respectively\cite{cteq}. The renormalization scale $\mu_R$ and factorization scale $\mu_F$ are
chosen to be $\mu=\mu_R=\mu_F$.
\begin{figure}[th]
  \centering
  \includegraphics[width=8cm]{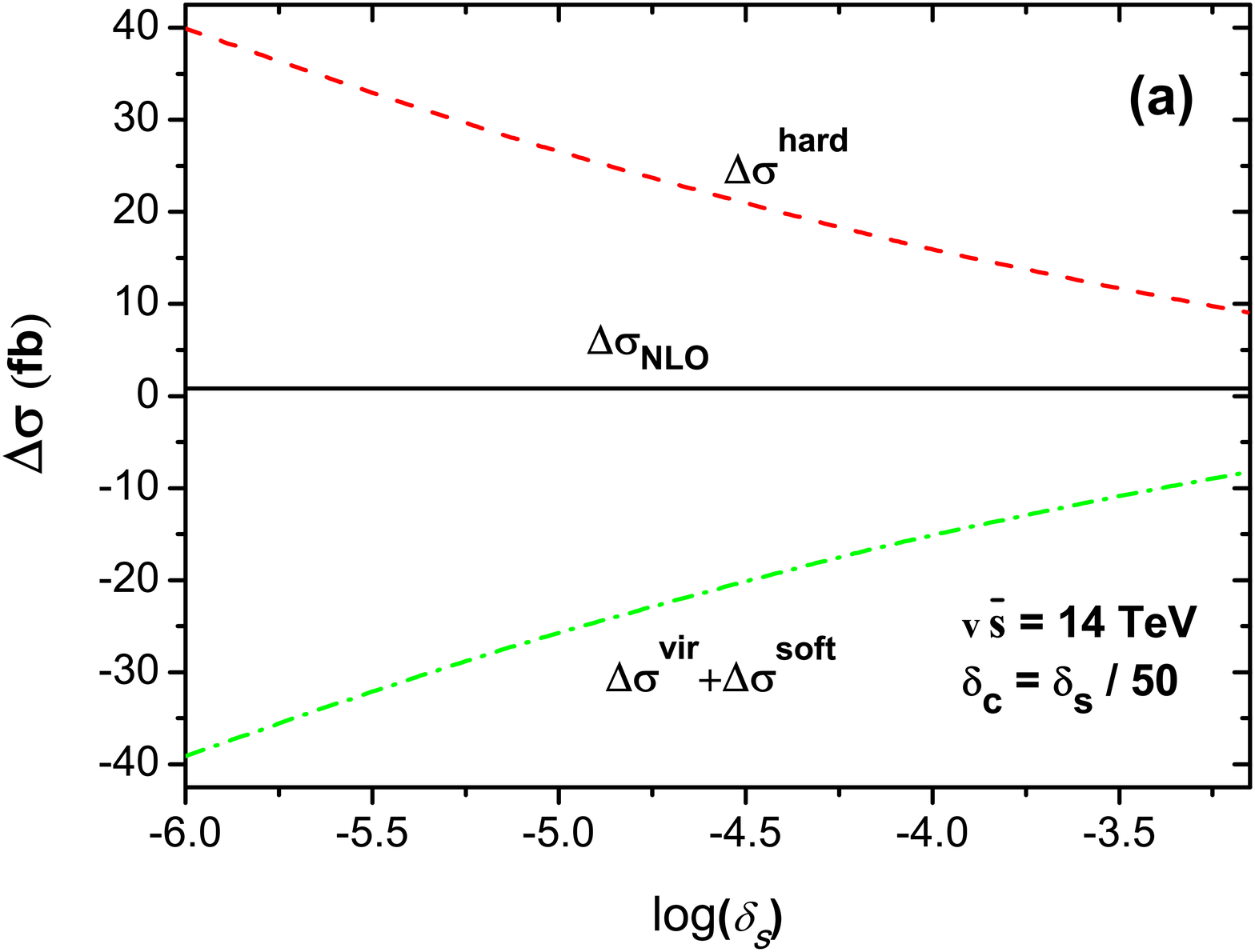}
  \includegraphics[width=8cm]{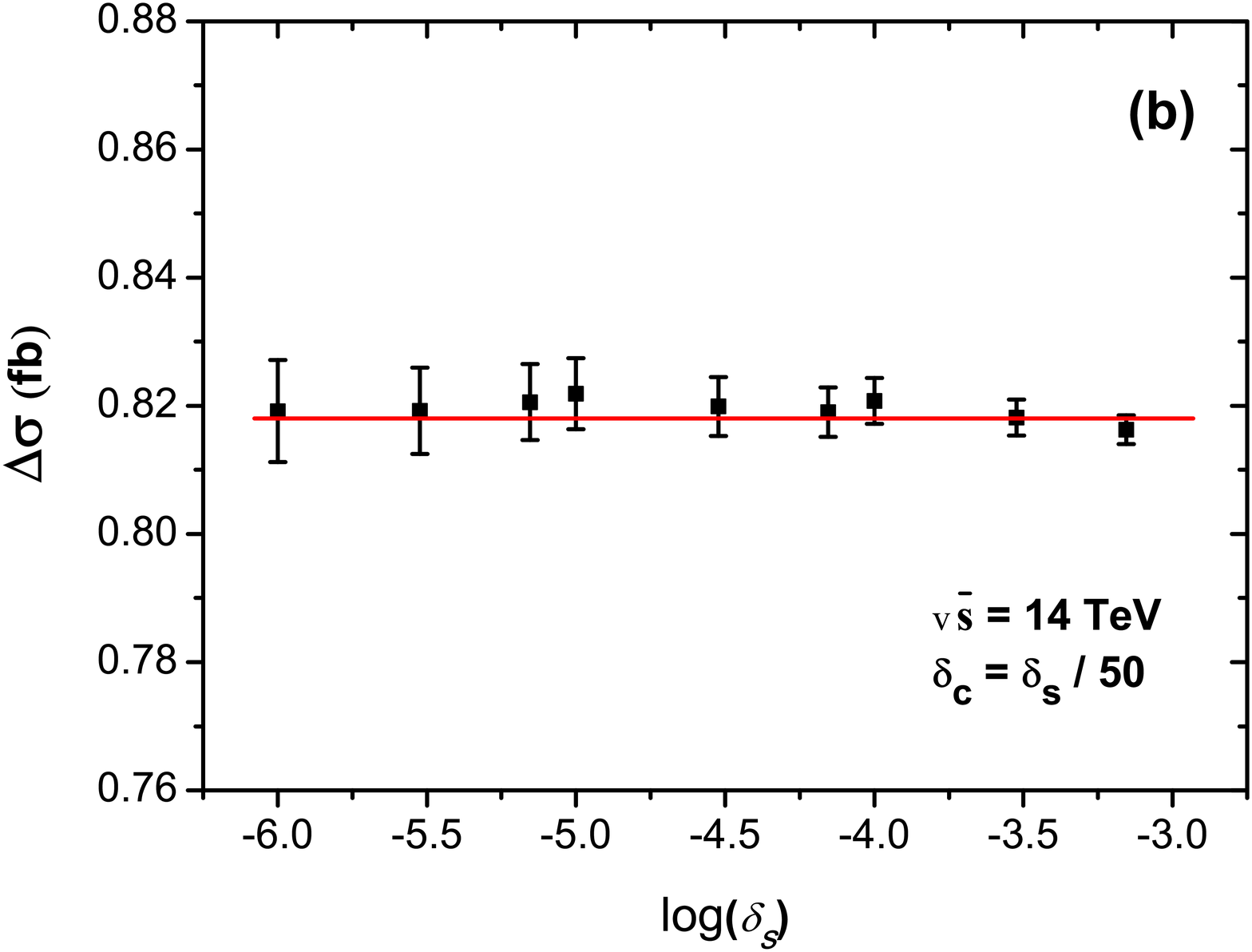} \vspace{-0.5cm}
  \caption{(a) The dependence of the NLO QCD corrections to $ pp \to HZW^+ $ on cutoffs $\delta_s$
  and $ \delta_c$ at 14 TeV LHC, where $\delta_s = 10^{-4}$ and $ \delta_c = \delta_s /50 $;
  (b) The amplified curve marked with the calculation errors for $\Delta\sigma_{NLO}$ versus $\delta_s$.}\label{cutoff}
\end{figure}

Although the splitting of phase space depends on the cutoff parameters $\delta_s$ and $\delta_c$, the NLO QCD corrections should be independent of these unphysical parameters. In the TCPSS method\cite{TCPSS}, $\delta_c$ is inclined to be much smaller than $\delta_s$ to guarantee the accuracy in the numerical calculations. Here we assume the collinear cutoff parameter $\delta_c = \delta_s /50$ and the renormalization scale $\mu_0 = (m_H + m_Z + m_W)/2 $. From the left panel of Fig.\ref{cutoff}, we can see that the values of $\Delta\sigma_{hard}$ and $\Delta\sigma_{vir}+\Delta\sigma_{soft}$ change with the variation of the soft cutoff $\delta_s$. But the total correction $\Delta\sigma_{NLO}$ is independent of the $\delta_s$ within the reasonable calculation errors. On the right panel of Fig.\ref{cutoff}, we display the amplified curve of $\Delta\sigma_{NLO}$ versus $\delta_s$. Therefore, we can take $ \delta_s = 10^{-4} $ and $\delta_c=2\times10^{-5}$ in the following calculations.
\begin{figure}[th]
  \centering
  \includegraphics[width=8cm]{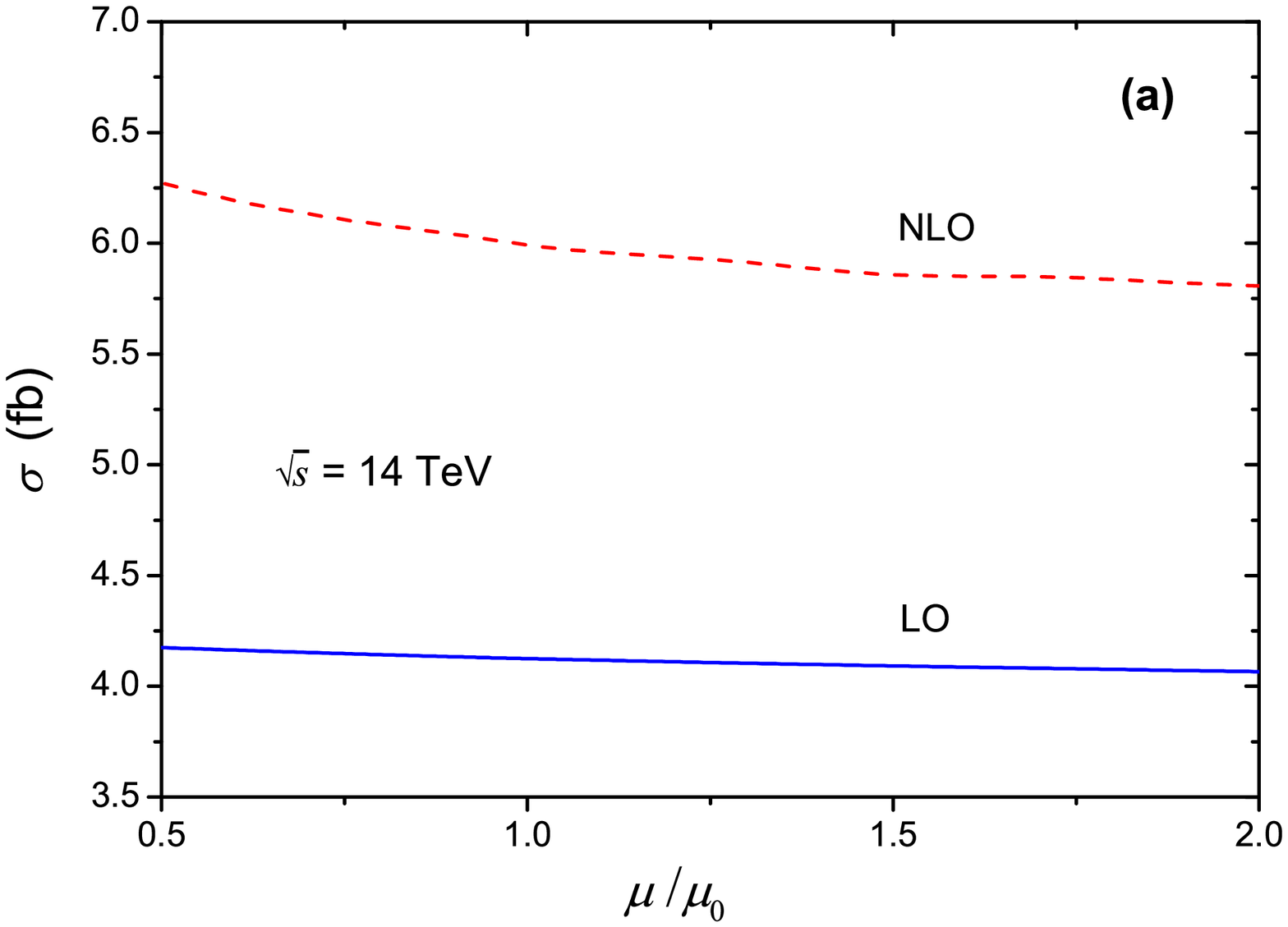}
  \includegraphics[width=8cm]{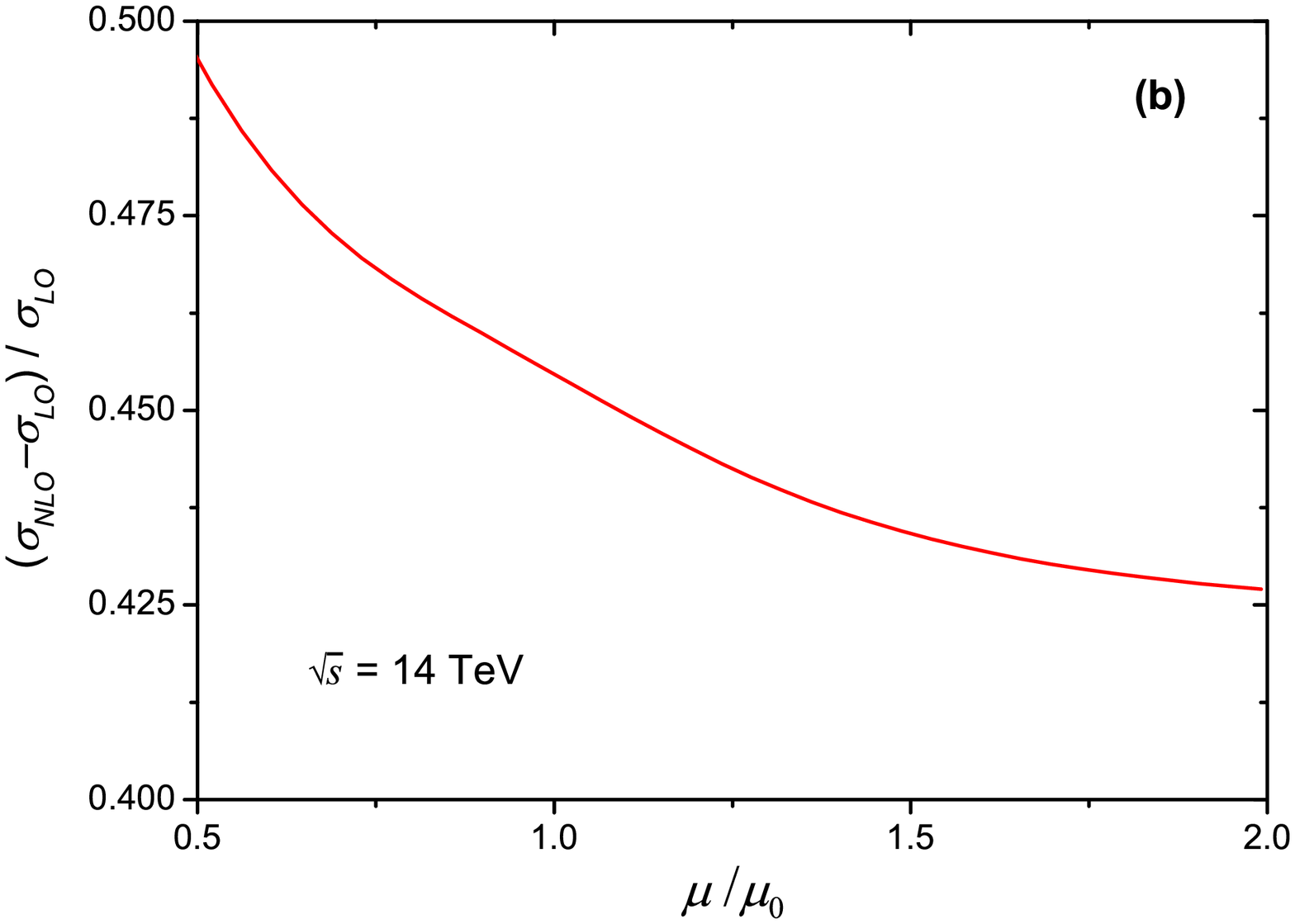}
  \caption{(a) The dependence of the LO and NLO QCD corrected cross sections for $ pp \to HZW $ process on
   the renormalization and factorization scale $ \mu $; (b) the relative correction of the total NLO corrected
   cross section to the LO cross section. }\label{total}
\end{figure}

We present the dependence of LO and NLO QCD corrected cross sections of $pp \to HZW^{\pm}$ production on the renormalization and factorization scale $\mu$ at 14 TeV LHC in Fig.\ref{total}. It can be seen that the LO and NLO QCD corrected cross sections can reach 4.2 fb and 6 fb respectively and the corresponding relative correction is 46\% when $\mu = \mu_0 = (m_H + m_Z + m_W)/2 $. If we vary the scale $\mu$ from $ \mu_0/2$ to $2\mu_0$, we find that the uncertainty of NLO QCD cross section $_{-2.96}^{+4.51}\%$ is larger than the one of LO cross section $_{-1.42}^{+1.17}\%$. The reason is that the LO process of $pp \to HZW^{\pm}$ is induced by the pure electroweak interactions and is not sensitive to the change of renormalization scale.

Due to the unbalance of PDF, the cross section of $pp \to HZW^{+}$ is larger than the one of $pp \to HZW^{-}$. This asymmetry can be measured through the leptonic decay of $W$ boson by examining the difference between event numbers with one lepton and those with one anti-lepton in the signal. So we can define the charge asymmetry in the process of $pp \to HZW^{\pm}$ as follows:
\begin{eqnarray}
A_c &=& \frac{N(HZW^+)-N(HZW^-)}{N(HZW^+)+N(HZW^-)}.
\end{eqnarray}
where $N$ is the event number of final states $HZW^+$($HZW^-$). In Tab.\ref{tab1}, we display the charge asymmetry for different renormalization and factorization scale $\mu$. We can see that the values of $A_c$ have the weak dependence on the scale $\mu$, due to the cancellation between numerator and denominator. Compared with the LO prediction of $A_c$, the NLO QCD correction will slightly reduce the value of $A_c$ to 32.94\% when $\mu=\mu_0$.

\begin{table}[t]
\begin{tabular}{p{2cm}<{\centering} p{2cm}<{\centering} p{2cm}<{\centering} p{2cm}<{\centering}}
\hline
  $A_c$(\%) & $\mu_0/2$ & $\mu_0$ & $2\mu_0$ \\
\hline
  LO & 33.38 & 33.43 & 33.50 \\
  NLO & 32.70 & 32.94 & 33.46 \\
\hline
\end{tabular}
\caption{The charge asymmetry of $W^{\pm}$ in the production of $pp\to HZW^{\pm}$ at renormalization and factorization scale $\mu=\mu_0/2,\mu_0,2\mu_0$ at 14 TeV LHC.\label{tab1}}
\end{table}

\begin{figure}[th]
  \centering
  \includegraphics[width=8cm,height=6cm]{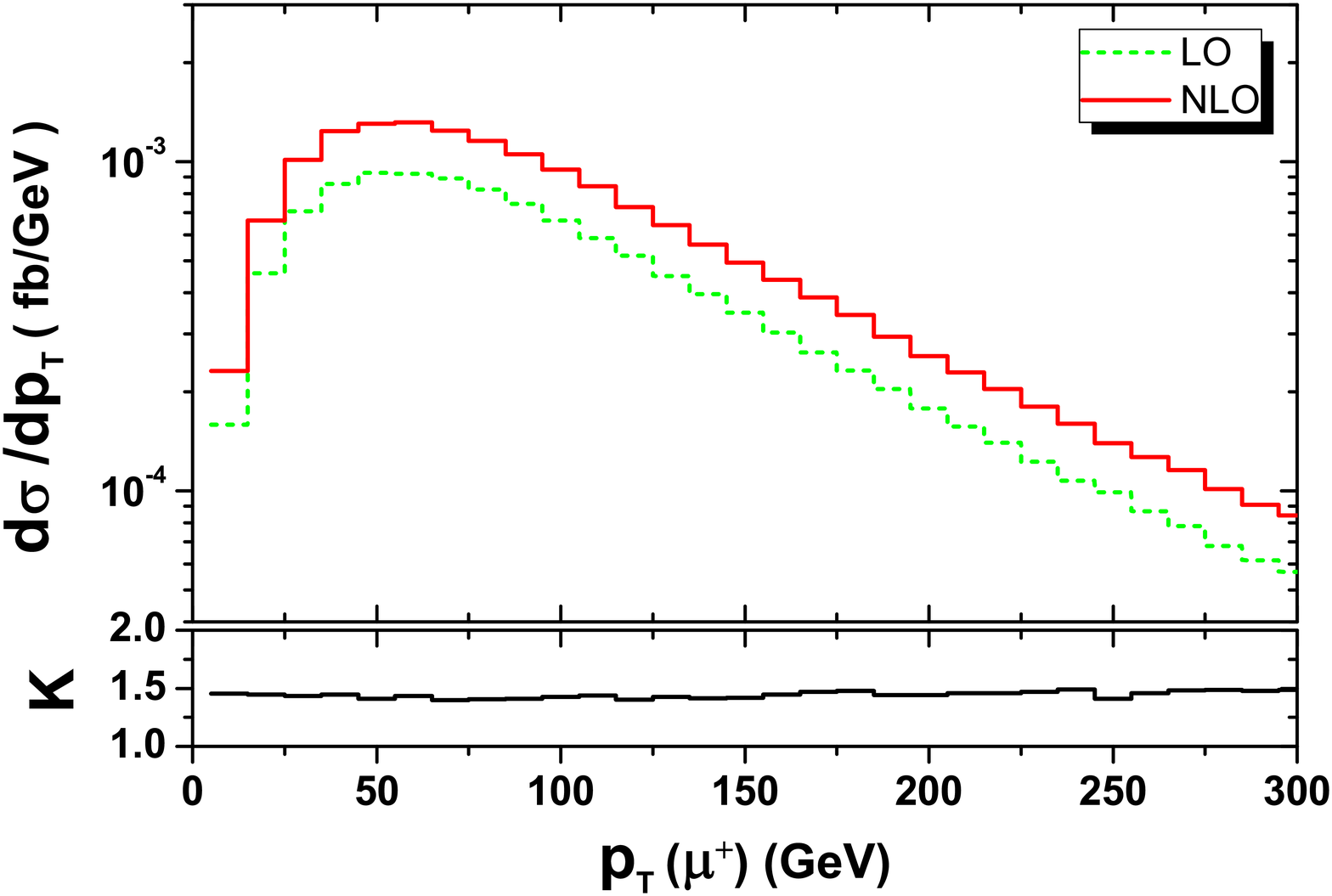}
  \includegraphics[width=8cm,height=6.2cm]{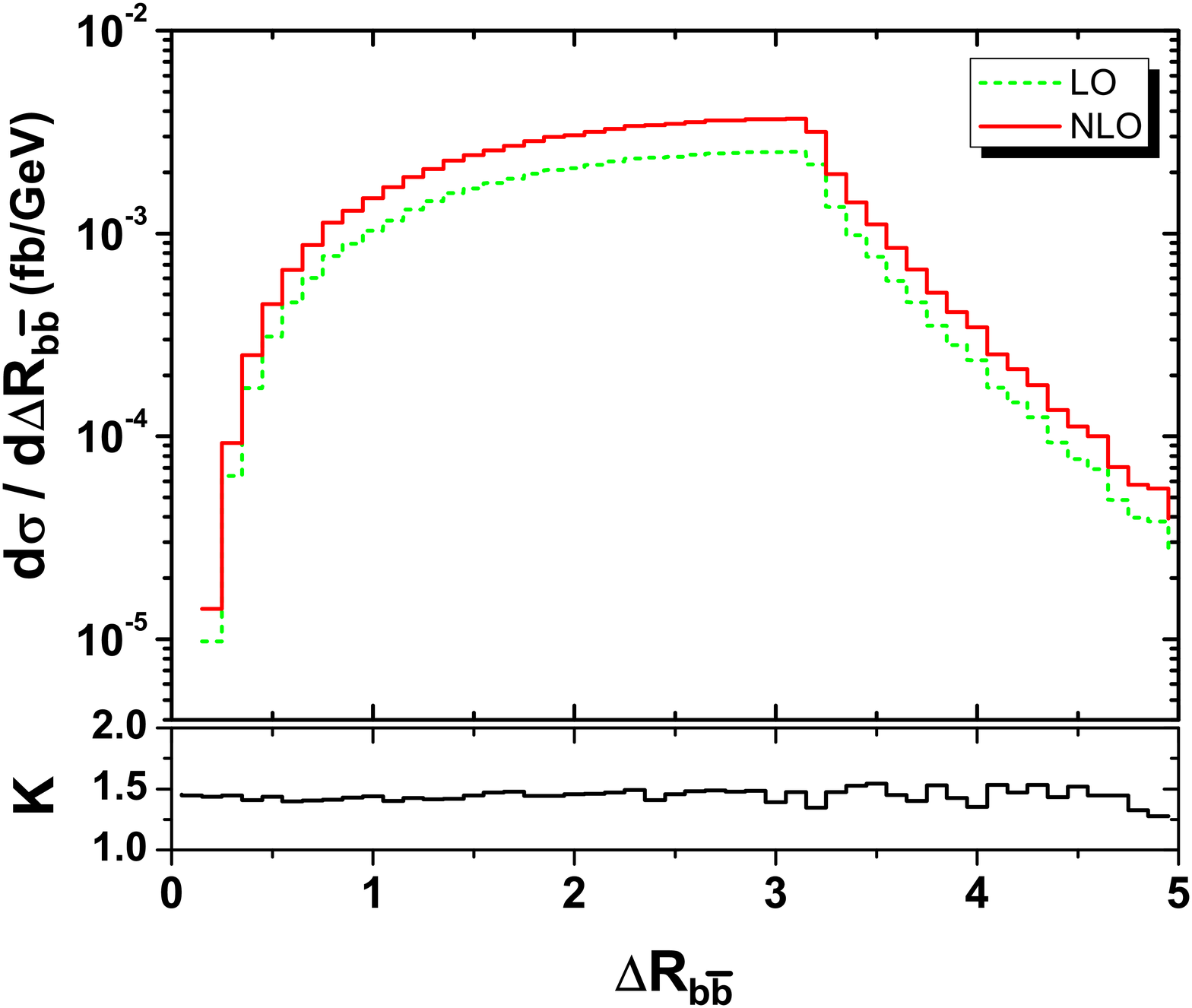}
  \caption{The LO and NLO QCD corrected transverse momentum distributions of $\mu^+$ from $W^+$ decays( $p^{\mu^+}_T$ ) and the separation between the two $b$-jets from Higgs boson decays ($\Delta R_{bb}$) and the corresponding $K$-factors in the process $pp\to HZW^{+}$ at 14 TeV LHC.}\label{distribution}
\end{figure}

In Fig.\ref{distribution}, we study the LO and NLO QCD corrected transverse momentum distributions of $\mu^+$ from $W^+$ decays( $p^{\mu^+}_T$ ) and the separation between the two $b$-jets from Higgs boson decays ( $\Delta R_{bb} \equiv \sqrt{(\Delta \phi)^2 + (\Delta \eta)^2}$ ) in the process of $pp\to HZW^{+}$ at 14 TeV LHC. In the calculations, we use the branching ratios of $Br(H \to b\bar{b})=57.8\%$\cite{brhbb} and $Br(W^+ \to \mu^+\nu_\mu)=10.57\%$\cite{pdg}. From Fig.\ref{distribution}, we can see that the NLO QCD correction can greatly enhance the LO differential cross section. The distribution of $p^{\mu^+}_T$ peaks around 40 GeV($\sim m_W/2$) and the two b-jets are incline to fly back-to-back since they come from the Higgs boson.

The main backgrounds of the signal $pp \to HZW^{\pm} \to 2b+3l+E^{miss}_{T}$ are: (i) $pp \to t\bar{t} \to 2b+2\ell+E^{miss}_T$; (ii) $pp \to Zb\bar{b} \to 2b+2\ell$; (iii) $pp \to t\bar{t}W^{\pm} \to 2b+3\ell+E^{miss}_T$. For (i)-(ii), they may resemble the signal when an ISR/FSR jet is mis-identified as a lepton. We notice that these backgrounds are much larger than our signal so that the observation of the signal may be challenge at the high luminosity LHC(HL-LHC). However, the following kinematic features may be helpful to improve the observability: since the two $b$-jets in the backgrounds are from the top quark and anti-top quark, a narrow window of the invariant mass of $b\bar{b}$ around 125 GeV can be used to greatly reduce the above backgrounds. Besides, the fake lepton from ISR/FSR jet is usually not energetic, so the veto of the third soft lepton will further suppress (i) and (ii) backgrounds. It should be noted that the background (ii) can also be removed by imposing the cut of the transverse mass of lepton and missing energy, because the fake lepton does not come from the $W$ boson decay. On the other hand, the missing energy in the backgrounds (i) and (iii) are larger than that in the signal, due to the leptonic decay top quark pair. Thus, we can also reduce these two backgrounds by requiring a small missing energy. Of course, the results of these cuts strongly depend on the detector simulation. However, since the realistic detector configures of the HL-LHC are still not available, we expect our analysis can be improved by optimizing signal extraction strategies and better understanding of the backgrounds uncertainties through the dedicated analysis of the experimental collaborations at HL-LHC.

\section{ Conclusion }

In this paper, we performed a complete NLO QCD corrections calculation for $pp\to HZW^{\pm}$ at 14 TeV LHC. We found that the NLO QCD corrections can enhance the LO cross section and the relative correction can reach about 45\%. We investigated the dependence of the LO and NLO corrected cross sections on the renormalization and factorization scale and found that the scale uncertainty of LO and NLO QCD cross section are $_{-1.42}^{+1.17}\%$ and $_{-2.96}^{+4.51}\%$ respectively, when the scale $\mu$ changes from $ \mu_0/2$ to $2\mu_0$. We also studied the kinematic distributions of the final state, which will be helpful to select the $HZW^{\pm}$ events at 14 TeV LHC.

\section*{Acknowledgement}
We appreciate the helpful discussions with Lei Wu. Ning Liu would
like to thank Dr Archil Kobakhidze for his warm hospitality in
Sydney node of CoEPP in Australia. This work is supported by the
National Natural Science Foundation of China (NNSFC) under grant
Nos.11305049 and 11275057, by Specialized Research Fund for the Doctoral Program of Higher Education
under Grant No.20134104120002, and by the Startup Foundation for Doctors
of Henan Normal University under contract No.11112.

\end{document}